\documentclass[aps, prl, twocolumn, superscriptaddress]{revtex4-2}
\usepackage{bm, amsmath, amsfonts, amssymb, ascmac, mathtools, braket}
\usepackage{multirow}
\usepackage{graphicx}
\usepackage{float, color, xcolor}
\usepackage{physics}

\usepackage{bbm}
\usepackage{tabularx}
\usepackage{enumerate}
\usepackage{comment}

\usepackage[
pagebackref=false,
colorlinks=true,
linkcolor=blue,
urlcolor=blue,
filecolor=black,
citecolor=red,
pdfstartview=FitV,
pdftitle={},
pdfauthor={},
pdfsubject={},
pdfkeywords={},
pdfpagemode=None,
bookmarksopen=true
]{hyperref}

\newcommand{\ii}{\text{i}}

\begin{document}

\title{Non-Hermitian Topology in Hermitian Topological Matter}

\author{Shu Hamanaka}
\email{hamanaka.shu.45p@st.kyoto-u.ac.jp}
\affiliation{Department of Physics, Kyoto University, Kyoto 606-8502, Japan}
\affiliation{Institute for Theoretical Physics, ETH Zurich, 8093 Zurich, Switzerland}

\author{Tsuneya Yoshida}
\email{yoshida.tsuneya.2z@kyoto-u.ac.jp}
\affiliation{Department of Physics, Kyoto University, Kyoto 606-8502, Japan}
\affiliation{Institute for Theoretical Physics, ETH Zurich, 8093 Zurich, Switzerland}

\author{Kohei Kawabata}
\email{kawabata@issp.u-tokyo.ac.jp}
\affiliation{Institute for Solid State Physics, University of Tokyo, Kashiwa, Chiba 277-8581, Japan}

\date{\today}

\begin{abstract}
Non-Hermiticity gives rise to distinctive topological phenomena absent in Hermitian systems.
However, connection between such intrinsic non-Hermitian topology and Hermitian topology has remained largely elusive.
Here, considering the bulk and boundary as an environment and system, respectively, we demonstrate that anomalous boundary states in Hermitian topological insulators exhibit non-Hermitian topology.
We study the self-energy capturing the particle exchange between the bulk and boundary, and show that it detects Hermitian topology in the bulk and induces non-Hermitian topology at the boundary. 
As an illustrative example, we reveal non-Hermitian topology and concomitant skin effect inherently embedded within chiral edge states of Chern insulators. 
We also identify the emergence of hinge states within effective non-Hermitian Hamiltonians at surfaces of three-dimensional topological insulators.
Furthermore, we comprehensively classify our correspondence across all the tenfold symmetry classes of topological insulators and superconductors.
Our work uncovers hidden connection between Hermitian and non-Hermitian topology, and provides an approach to identifying non-Hermitian topology in quantum matter.
\end{abstract}

\maketitle

Topological phases of matter are a central topic in modern condensed matter physics~\cite{HK-review, QZ-review}. 
Gapped phases of noninteracting fermions are systematically classified by the tenfold fundamental symmetry~\cite{AZ-97}, culminating in the periodic table of topological insulators and superconductors~\cite{Schnyder-08, *Ryu-10, Kitaev-09, CTSR-review}.
A hallmark of topological insulators is the bulk-boundary correspondence:
the nontrivial bulk topology yields anomalous gapless states at boundaries.

Beyond the Hermitian regime, topological characterization of non-Hermitian systems has recently attracted growing interest~\cite{Rudner-09, Sato-11, *Esaki-11, Hu-11, Schomerus-13, Longhi-15, Lee-16, Leykam-17, Xu-17, Shen-18, Takata-18, MartinezAlvarez-18, Gong-18, *Kawabata-19, YW-18-SSH, *YSW-18-Chern, Kunst-18, McDonald-18, Lee-Thomale-19, Liu-19, Lee-Li-Gong-19, KSUS-19, ZL-19, Herviou-19, Zirnstein-19, Borgnia-19, Yokomizo-19, Chang-20, Wanjura-20, Zhang-20, OKSS-20, Yang-20, Denner-21, *Denner-23JPhysMater, Okugawa-20, KSS-20, Fu-21, KSR-21, Zhang-22, Sun-21, Kim-21, Tai-23, Nakamura-24, Wang-24, Nakai-24, BBK-review, Okuma-Sato-review}.
Non-Hermiticity arises from the exchange of particles and energy with the environment~\cite{Konotop-review, Christodoulides-review}.
Even within closed systems, non-Hermiticity of self-energy characterizes finite-lifetime quasiparticles~\cite{Kozii-17, *Papaj-19, *Shen-18QO, Yoshida-18, Hirsbrunner-19, McClarty-19, Bergholtz-19, Michishita-20, Okuma-21Green, Lehmann-21, Cayao-23}.
Non-Hermiticity enables a unique gap structure of complex spectra---point gap---and concomitant topological phases that have no analogs in Hermitian systems~\cite{Gong-18, KSUS-19}.
As the bulk-boundary correspondence in non-Hermitian systems, nontrivial point-gap topology leads to skin effect~\cite{Zhang-20, OKSS-20} and anomalous boundary states~\cite{OKSS-20, Denner-21, *Denner-23JPhysMater, KSR-21, Nakamura-24}.
Such intrinsic non-Hermitian topological phenomena have been experimentally realized across various open classical and quantum systems~\cite{Poli-15, Zeuner-15, Zhen-15, Weimann-17, Xiao-17, St-Jean-17, Parto-17, Bahari-17, Zhao-18, Zhou-18, Harari-18, *Bandres-18, Cerjan-19, Zhao-19, Brandenbourger-19-skin-exp, *Ghatak-19-skin-exp, Helbig-19-skin-exp, *Hofmann-19-skin-exp, Xiao-19-skin-exp, Weidemann-20-skin-exp, Gou-20, *Liang-22, Palacios-21, Zhang-21, Wang-23mech, Shen-23, *Shen-24}, including photonic~\cite{Poli-15, Zeuner-15, Zhen-15, Weimann-17, St-Jean-17, Parto-17, Bahari-17, Zhao-18, Zhou-18, Harari-18, *Bandres-18, Cerjan-19, Zhao-19, Weidemann-20-skin-exp}, mechanical~\cite{Brandenbourger-19-skin-exp, *Ghatak-19-skin-exp, Wang-23mech}, and electrical~\cite{Helbig-19-skin-exp, *Hofmann-19-skin-exp} systems, along with single photons~\cite{Xiao-17, Xiao-19-skin-exp}, ultracold atoms~\cite{Gou-20, *Liang-22}, and digital quantum processors~\cite{Shen-23, *Shen-24}.

Notably, the topological classification suggests the correspondence of Hermitian topology in $d$ dimensions and non-Hermitian topology in $d-1$ dimensions~\cite{KSUS-19, JYLee-19, Bessho-21}.
In line with this correspondence, non-Hermitian perturbations can induce point gaps for anomalous boundary states in topological insulators~\cite{Li-22, Zhu-22, *Zhu-23, Ou-23, Ma-24, Schindler-23, Nakamura-23}, as observed in recent photonic~\cite{Liu-24} and phononic~\cite{Wu-23} experiments.
However, the direct connection between Hermitian and non-Hermitian topology has remained elusive, given that non-Hermiticity is merely added as an external perturbation.
Consequently, the fundamental mechanism underlying the correspondence between the Hermitian bulk and non-Hermitian boundary has been unclear.

\begin{figure}[t]
\centering
\includegraphics[width=0.8\linewidth]{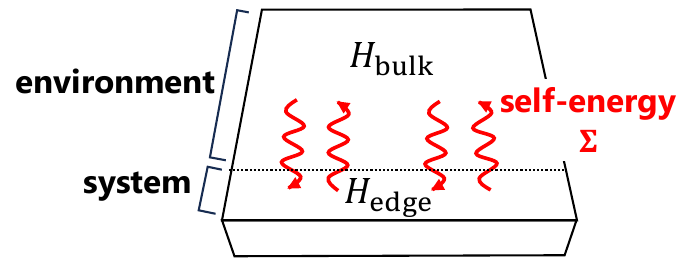} 
\caption{Effective non-Hermitian Hamiltonian $H_{\rm edge} + \Sigma$ at the edge of a Hermitian Hamiltonian.
The self-energy $\Sigma$ describes the particle exchange between the bulk $H_{\rm bulk}$ and edge $H_{\rm edge}$, yielding non-Hermitian topology.}	
    \label{fig: setup}
\end{figure}

In this Letter, we reveal non-Hermitian topology inherently embedded within Hermitian topological matter.
Regarding the bulk and boundary in Hermitian topological insulators as an environment and system, respectively, we show that effective boundary Hamiltonians exhibit non-Hermitian topology (Fig.~\ref{fig: setup}).
As an illustrative example, we demonstrate the non-Hermitian skin effect in chiral edge states of Chern insulators.
Furthermore, we systematically classify our correspondence across all the tenfold symmetry classes of topological insulators and superconductors (Table~\ref{tab:periodictable-h}).
Our work uncovers hidden microscopic connection between Hermitian and non-Hermitian topology.

\textit{Non-Hermitian Hamiltonians in closed systems}.---Our central idea involves conceptually dividing the bulk and boundary of a closed system into an {\it{environment}} and {\it{system}}, respectively (Fig.~\ref{fig: setup}). 
The single-particle Hamiltonian of the entire system reads
\begin{equation}
    H = \begin{pmatrix}
        H_{\rm bulk} & T \\
        T^{\dag} & H_{\rm edge}
    \end{pmatrix},
        \label{eq: Ham-tot}
\end{equation}
where $H_{\rm bulk}$ ($H_{\rm edge}$) is the Hamiltonian in the bulk (at the boundary), and $T$ denotes the coupling between the bulk and boundary.
From the original Schr\"odinger equation $H\,( \ket{\psi}_{\rm bulk} \,\ket{\psi}_{\rm edge} )^T = \left( E+\ii \eta \right) ( \ket{\psi}_{\rm bulk}\,\ket{\psi}_{\rm edge} )^T$ with an infinitesimal number $\eta > 0$ reflecting causality, we project the boundary degree of freedom and derive the effective Hamiltonian $H_{\mathrm{eff}} \left( E \right) = H_{\mathrm{edge}} + \Sigma \left( E \right)$ with the self-energy~\cite{Datta-textbook, Moiseyev-textbook, supplement}
\begin{equation}
    \Sigma \left( E \right) \coloneqq T^{\dag} \left( E + \ii \eta - H_{\rm bulk} \right)^{-1} T.
        \label{eq: self-energy}
\end{equation}
The self-energy $\Sigma \left( E \right)$ captures the continuous exchange of particles between the 
bulk and boundary for given energy $E \in \mathbb{R}$.
Consequently, $\Sigma \left( E \right)$ acquires non-Hermiticity, describing finite lifetimes of quasiparticles that escape from the boundary to the bulk.
Non-Hermiticity of $\Sigma \left( E \right)$ emerges through the sufficiently strong bulk-boundary coupling $T$, which cannot be captured by the Lindblad master equation due to its inherent weak coupling assumption~\cite{GKS-76, Lindblad-76, Breuer-textbook}.

Crucially, topology of the original Hermitian Hamiltonian $H$ should leave an imprint on that of the effective non-Hermitian Hamiltonian $H_{\rm eff} \left( E \right)$. 
For example, when $H$ is a quantum Hall (Chern) insulator, $H_{\rm eff} \left( E \right)$ effectively describes chiral edge states for $E$ within an energy gap. 
Intuitively, chirality or nonreciprocity of the anomalous boundary states should yield non-Hermitian topology, as can be seen in the celebrated Hatano-Nelson model~\cite{Hatano-Nelson-96, *Hatano-Nelson-97}.
In this Letter, we substantiate this intuition and demonstrate that the self-energy bridges Hermitian bulk topology and non-Hermitian boundary topology.
Below, for fixed $E$ (i.e., Markov approximation~\cite{Michishita-20, Hatano-21}),
we study non-Hermitian topology of $H_{\rm eff} \left( E \right)$ in prototypical topological insulators.

\textit{Non-Hermitian topology in the Su-Schrieffer-Heeger model}.---We begin with a one-dimensional topological
insulator, Su-Schrieffer-Heeger (SSH) model~\cite{SSH-79}. 
The Bloch Hamiltonian reads
\begin{equation}
    \label{eq:ssh}
    H_{\mathrm{SSH}} \left( k \right) = \left( v + t \cos{k} \right) \sigma_x + \left( t \sin{k} \right) \sigma_y,
\end{equation}
where $\sigma_i$'s ($i=x, y, z$) denote Pauli matrices, and 
$v > 0$ and $t > 0$ are
the intracell and intercell hopping amplitudes, respectively.

\begin{figure}[t]
\centering
\includegraphics[width=\linewidth]{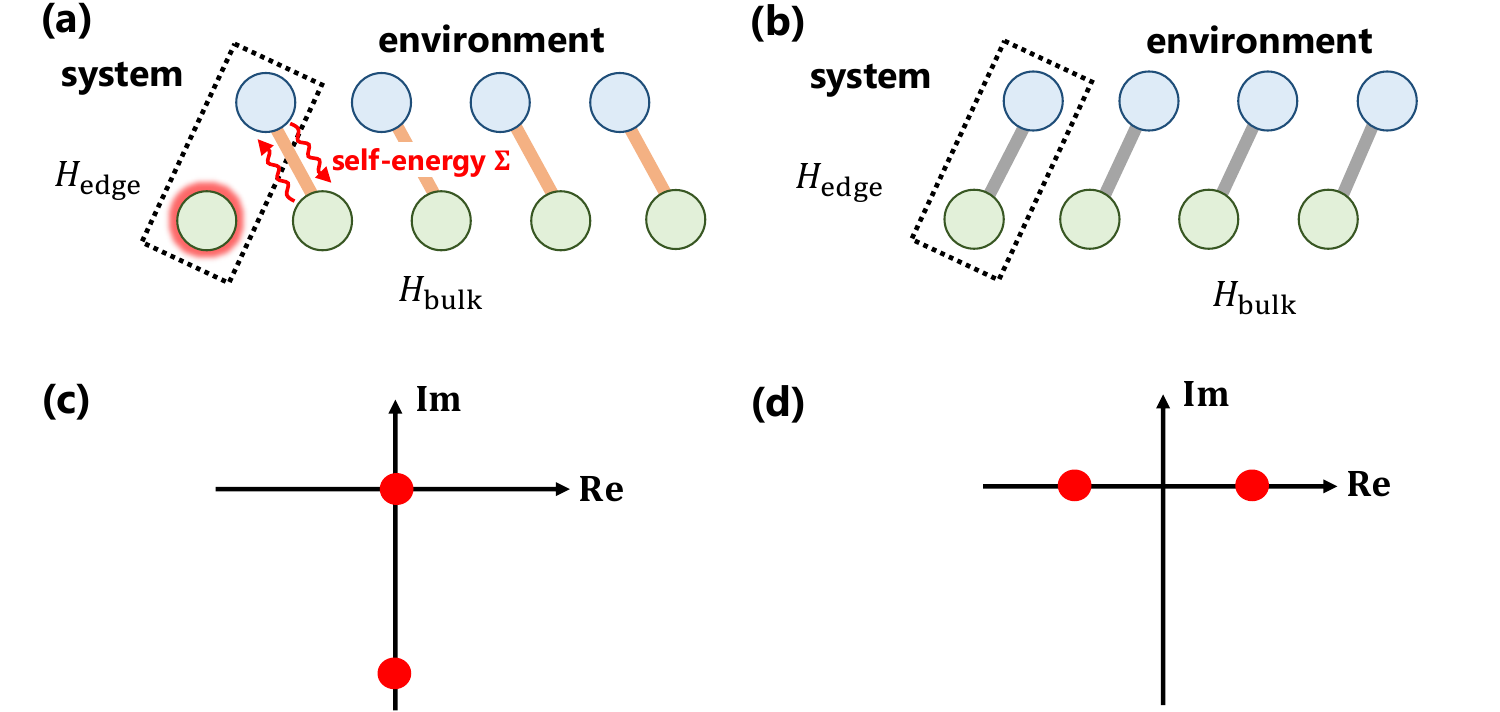} 
\caption{Non-Hermitian topology in the Su-Schrieffer-Heeger model.
(a, b)~Correspondence of Hermitian and non-Hermitian topology for the (a)~nontrivial and (b)~trivial phases.
(c, d)~Complex eigenvalues of the effective non-Hermitian Hamiltonian at the edge for the (c)~nontrivial and (d)~trivial phases.}	
    \label{fig: ssh}
\end{figure}

The SSH model respects chiral symmetry $\sigma_z H_{\rm SSH} \left( k \right) \sigma_z = - H_{\rm SSH} \left( k \right)$ and belongs to class AIII, characterized by the integer topological invariant~\cite{CTSR-review}.
In the topological phase ($v/t <1$), the bulk topology leads to the emergence of an edge state $\ket{\psi_0} \propto 
\sum_{x} \left( -v/t \right)^x \ket{x} \otimes \left( 1~0 \right)^T$ with zero energy $E=0$ under the open boundary conditions.
When we regard the unit cell at the left edge as a system and the remaining portion an environment [Fig.~\ref{fig: ssh}~(a,~b)], quasiparticles at the edge are coupled with the bulk and can escape into the bulk.
This decaying property is quantified by the self-energy in Eq.~(\ref{eq: self-energy}), calculated as~\cite{supplement}
\begin{equation}
    \Sigma \left( E \right) =
      \begin{pmatrix}
        0 & 0 \\
        0 & -\ii \pi \left( t^2-v^2 \right) \delta \left( E \right) \theta \left( t-v \right)
    \end{pmatrix}.
        \label{eq: SSH - selfenergy}
\end{equation}
The complex eigenvalues of the effective non-Hermitian Hamiltonian $H_{\rm eff} \left( 0 \right) = v \sigma_x + \Sigma \left( 0 \right)$ are obtained as $0$ and $-\ii\infty$, 
the former of which corresponds to a topologically stable zero state with an infinite lifetime and the latter of which an unstable zero state with a vanishing lifetime.
In the trivial phase ($v/t > 1$), by contrast, no eigenstates appear at $E=0$, and hence the self-energy vanishes $\Sigma \left( 0 \right) = 0$.

The topological nature of the zero-energy edge state causes non-Hermitian topology of the effective Hamiltonian $H_{\rm eff} \left( 0 \right)$.
In the topological phase, the two single-particle eigenenergies are gapped with a reference energy on the imaginary axis [Fig.~\ref{fig: ssh}\,(c)], i.e., point gap~\cite{Gong-18, KSUS-19}.
Inheriting from chiral symmetry of the original SSH model $H_{\rm SSH} \left( k \right)$, the edge non-Hermitian Hamiltonian $H_{\rm eff} \left( 0 \right)$ also respects chiral symmetry $\sigma_z H_{\rm eff}^{\dag} \left( 0 \right) \sigma_z = - H_{\rm eff} \left( 0 \right)$.
Consequently, topologically-protected zero-energy states persist as long as the point gap is open.
Such persistence is ensured by point-gap topology, given by the zeroth Chern number of the Hermitian matrix $\ii H_{\rm eff} \left( 0 \right) \sigma_z$~\cite{KSUS-19}.
In fact, $H_{\rm eff} \left( 0 \right) = v\sigma_x + \Sigma \left( 0 \right)$ in Eq.~(\ref{eq: SSH - selfenergy}) exhibits the zeroth Chern number $1$ ($0$) in the topological (trivial) phase.
Thus, the self-energy $\Sigma \left( 0 \right)$ detects nontrivial Hermitian topology in the bulk and induces non-Hermitian topology at the edge.

\begin{figure}[t]
\centering
\includegraphics[width=\linewidth]{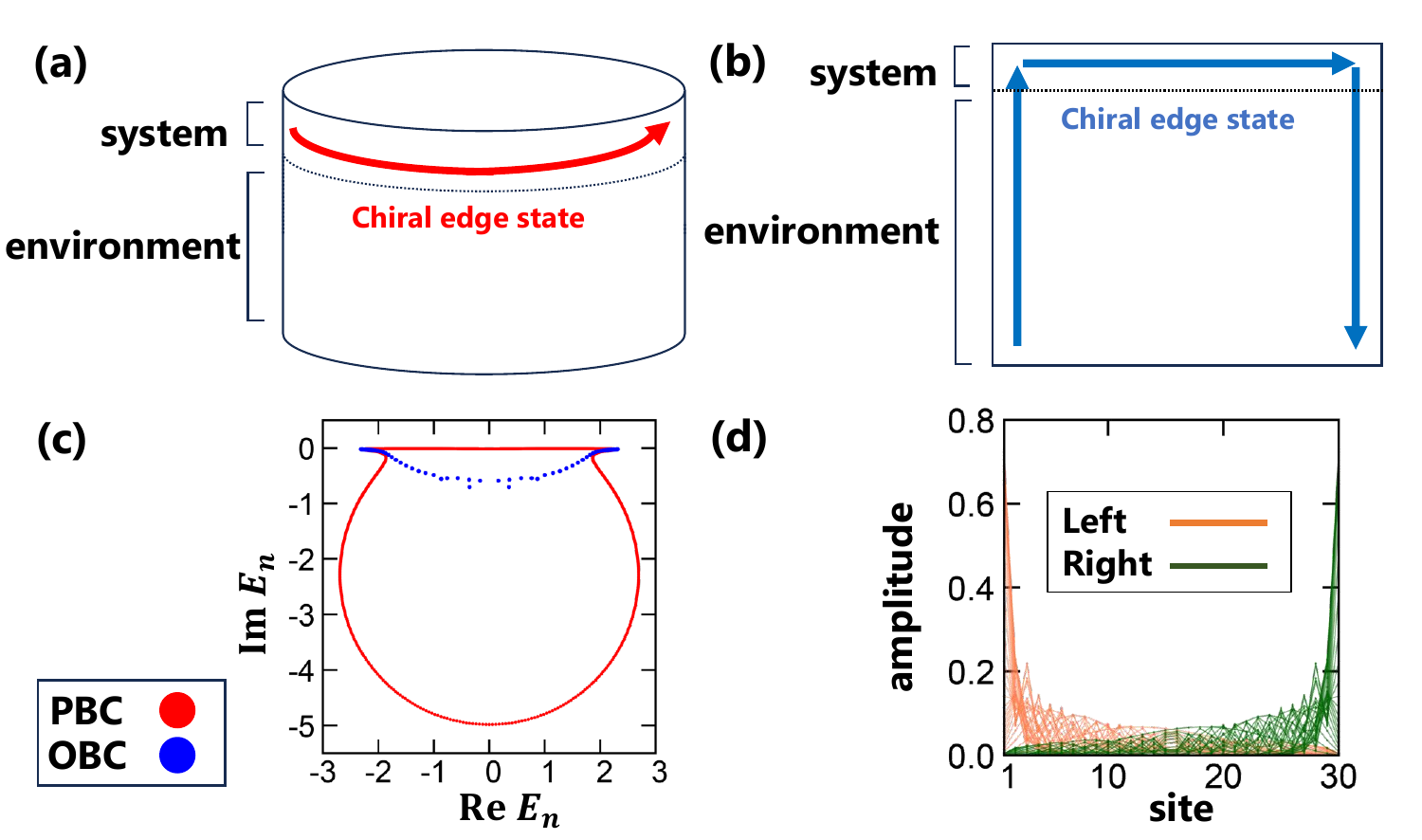} 
\caption{Non-Hermitian topology in the Chern insulator ($t=1.0$, $m=-1.3$, $E=0$).
The open boundary conditions (OBC) are imposed along the $x$ direction.
(a, b)~Chiral edge state as a non-Hermitian subsystem of the Chern insulator.
(c)~Complex spectrum of the effective non-Hermitian Hamiltonian under the periodic boundary conditions (PBC; red; $L_x = 30$, $L_y = 3000$) and OBC (blue; $L_x = L_y = 30$) along the $y$ direction. 
(d)~Collection of all right (orange) and left (green) skin states. 
The infinitesimal number $\eta$ is chosen as $\eta = 1/\sqrt{L_x}$~\cite{eta}.}	
\label{fig:qwz}
\end{figure}

\textit{Non-Hermitian topology in a Chern insulator}.---Next, we consider a Chern insulator on the square lattice described by~\cite{QWZ-06, *QHZ-08}
\begin{align}
    \label{eq:qwz}
    H_{\mathrm{Chern}} \left( \bm{k} \right) &= \left( t \sin{k_x} \right) \sigma_x + \left( t \sin{k_y} \right) \sigma_y \nonumber \\
    &\qquad + \left( m + t \cos{k_x}+t \cos{k_y} \right) \sigma_z
\end{align}
with $t,m \in \mathbb{R}$.
We regard the one-dimensional edge at $x=1$ as a system and the remaining bulk an environment,
showing that the effective edge Hamiltonian yields non-Hermitian topology and skin effect (Fig.~\ref{fig:qwz}).

This model $H_{\mathrm{Chern}} \left( \bm{k} \right)$ is characterized by the first Chern number, $C_1 = - \mathrm{sgn} \left( m/t \right)$ ($C_1 = 0$) for $\left| m/t \right| < 2$ ($\left| m/t \right| > 2$). 
Imposing the open and periodic boundary conditions along the $x$ and $y$ directions, respectively, we analytically obtain the self-energy~\cite{supplement}
\begin{equation}
    \label{eq:qwz-self}
    \Sigma \left( E, k_y \right) = \frac{t^2 - \left( m + t \cos{k_y} \right)^2}{2 \left( E+ \ii \eta-t\sin k_y \right)} 
    \left( \sigma_0 -\sigma_y \right).
\end{equation}
We calculate the complex eigenvalues of the effective non-Hermitian Hamiltonian $H_{\rm eff} \left( E, k_y \right) = \left( t\sin k_y \right) \sigma_y + \left( m+t\cos k_y \right) \sigma_z + \Sigma \left( E, k_y \right)$ [Fig.~\ref{fig:qwz}\,(c)].
They form a loop in the complex plane and host a point gap for reference energy $E_{0}$ inside the loop.
This loop structure intuitively originates from chirality of the anomalous boundary states, 
similarly to the Hatano-Nelson model~\cite{Hatano-Nelson-96, *Hatano-Nelson-97}.
The nontrivial point-gap topology is captured by the winding of the complex spectrum, 
\begin{equation}
    W_1 \coloneqq 
    - \oint \frac{dk}{2\pi\ii} \left( \frac{d}{dk} \log \det \left[ H \left( k \right) - E_0 \right] \right)
\end{equation}
for a Bloch Hamiltonian $H \left( k \right)$ in one dimension~\cite{Gong-18, KSUS-19}.
In fact, $H_{\rm eff} \left( E, k_y \right)$ with fixed $E$ exhibits $W_1 = - \mathrm{sgn} \left( m/t \right)$, leading to the correspondence $C_1 = W_1$ between the Hermitian bulk and non-Hermitian boundary~\cite{Schindler-23, Nakamura-23}.

The bulk-boundary correspondence for the nontrivial point-gap topology $W_1 \neq 0$ manifests itself as the non-Hermitian skin effect~\cite{Zhang-20, OKSS-20}.
We calculate the complex spectrum of $H_{\rm eff} \left( E, k_y \right)$ under the open boundary conditions along both $x$ and $y$ directions [Fig.~\ref{fig:qwz}\,(c)].
Clearly, the complex spectrum does not form a loop but an arc;
such extreme sensitivity to the boundary conditions is a signature of the non-Hermitian skin effect~\cite{Lee-16, YW-18-SSH, Kunst-18}.
Consistently, most of the right and left eigenstates of $H_{\rm eff}$ are localized at either corner of the square lattice [Fig.~\ref{fig:qwz}\,(d)].
Notably, $H_{\rm eff}$ incorporates nonlocal hopping in real space, accompanying the nonuniform generalized Brillouin zone~\cite{YW-18-SSH, Yokomizo-19, Yang-20, Tai-23}.
While Ref.~\cite{Eek-24} also discussed $E$-dependent skin effect in non-Hermitian models, we here find it within Hermitian models.
We also demonstrate the $\mathbb{Z}_2$ skin effect in a time-reversal-invariant topological insulator~\cite{BHZ-06, supplement}.

\begin{figure}[t]
\centering
\includegraphics[width=0.75\linewidth]{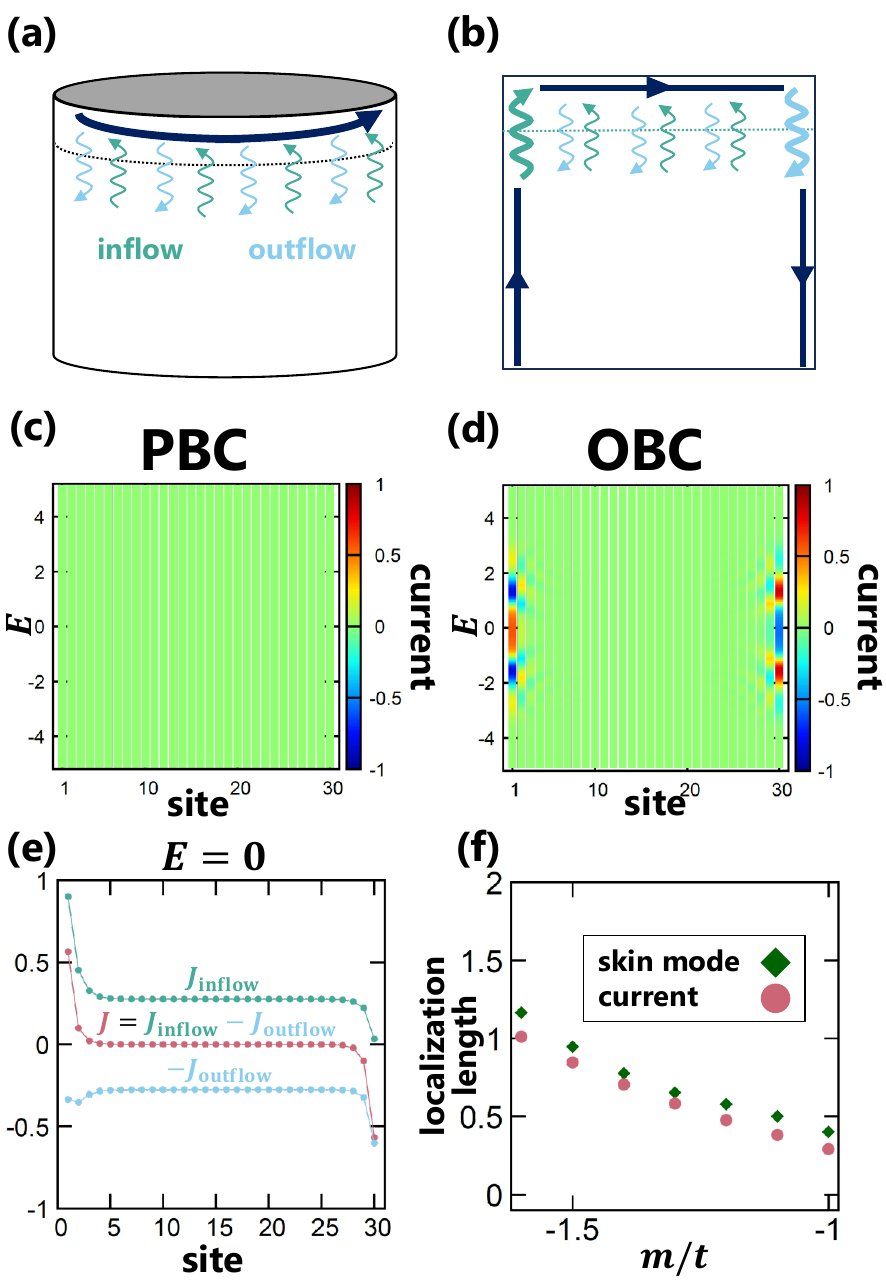} 
\caption{Localized current distribution due to the non-Hermitian skin effect.
The open boundary conditions (OBC) are imposed along the $x$ direction.
(a, b)~While the inflow and outflow balance (a)~under the periodic boundary conditions (PBC) along the $y$ direction, 
they do not (b)~under OBC.  
(c, d)~Terminal current for various energy $E$ under (c)~PBC and (d) OBC ($t=1.0$, $m=-1.3$, $L_x = L_y = 30$). 
(e)~Inflow $J_{\rm inflow}$ (green), net current $J = J_{\rm inflow}-J_{\rm outflow}$ (red), and outflow $-J_{\rm outflow}$ (blue) for $E=0$. 
(f)~Localization length of the current $J$ (red) and the most localized skin state (green) as functions of $m/t$, obtained from the scaling for the five sites near the boundary. 
The infinitesimal number $\eta$ is chosen as $\eta = 1/\sqrt{L_x}$~\cite{eta}.}
\label{fig:current}
\end{figure}

\textit{Skin current}.---We show that the non-Hermitian skin effect of the chiral edge states physically results in the localized current distribution (Fig.~\ref{fig:current}).
Imposing the open boundary conditions along the $x$ direction, we investigate the $E$-resolved local current~\cite{supplement,Datta-textbook,Datta-textbook-2005}
\begin{equation}
    \label{eq:current-formula}
    J \left( E \right) = -  \big[
   H_{\mathrm{edge}}, G_{\mathrm{edge}} \left( E \right)
   \big]
   - \big[
   H_{\mathrm{edge}}, G_{\mathrm{edge}} \left( E \right)
   \big]^\dag,
\end{equation}
with the edge Green's function $G_{\rm edge} \left( E \right) \coloneqq \left( E+\ii \eta - H_{\mathrm{eff}} \left( E \right) \right)^{-1}$.
Under the periodic boundary conditions along the $y$ direction, the inflow $J_{\rm inflow}$ and outflow $J_{\rm outflow}$ between the bulk and boundary balance, leading to the absence of the net current $J = J_{\rm inflow} - J_{\rm outflow} = 0$.
Under the open boundary conditions, by contrast, the nonzero net current $J$ arises around the corners.
This unique current distribution results from the chiral edge states.
Chirality of the edge states leads to a large inflow at one corner and a large outflow at the opposite corner.

The localized current distribution is a direct consequence of the corner skin effect and serves as an experimental probe of non-Hermitian topology in Hermitian Chern insulators.
We find that the localization length of the current density behaves consistently with that of the skin states [Fig.~\ref{fig:current}\,(f)].
A hallmark of the skin effect is the localization of right and left eigenstates at the opposite edges [see Fig.~\ref{fig:qwz}\,(d)].
From Eq.~(\ref{eq:current-formula}), a skin state with the localization length $\xi$ contributes to the the current density as $J_y \simeq j_1 e^{-y/\xi} - j_2 e^{-(L-y)/\xi}$ with $j_1, j_2 \in \mathbb{R}$, compatible with Fig.~\ref{fig:current}\,(e, f).
No local current arises if both right and left eigenstates of $H_{\rm eff} \left( E \right)$ are localized at the same boundary, as opposed to the skin states. 
Similar transport signatures of non-Hermitian topology were explored using scattering theory~\cite{Luis-20, Franca-22, Ochkan-24}.
We here elucidate the direct microscopic origin of these transport phenomena within Hermitian topological matter.

\begin{figure}[t]
\centering
\includegraphics[width=\linewidth]{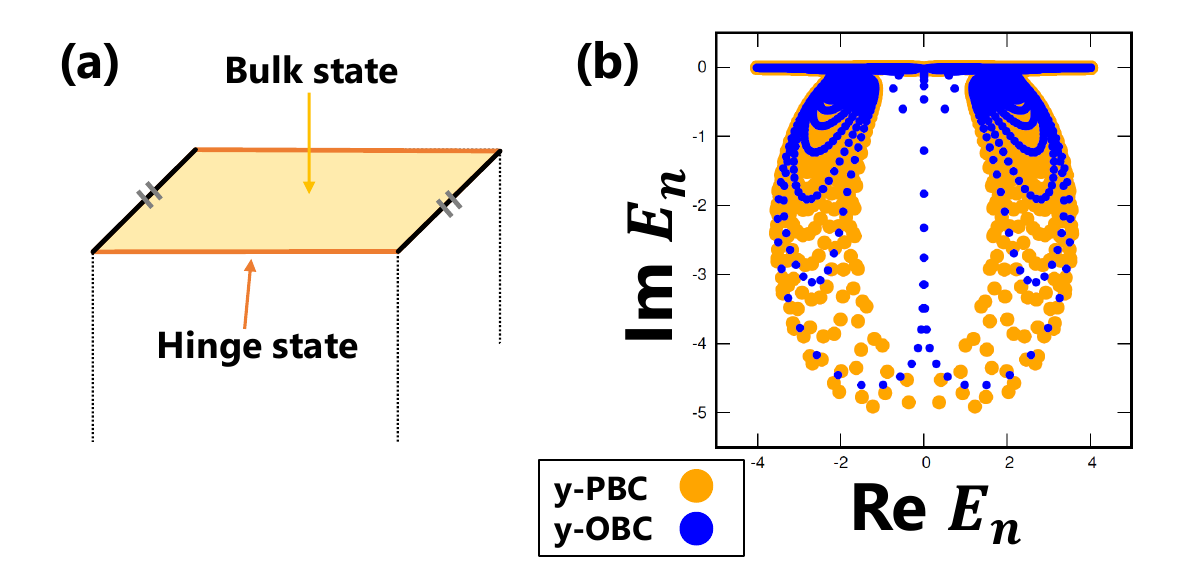} 
\caption{Non-Hermitian topology in a three-dimensional topological insulator.
(a)~Hinge state in an effective non-Hermitian Hamiltonian at a surface.
(b)~Complex spectrum of the effective non-Hermitian Hamiltonian under the periodic boundary conditions along the $y$ direction ($y$-PBC; orange; $L_x = L_y = 200$, $L_z = 30$) and open boundary conditions ($y$-OBC; blue; $L_x = 200$, $L_y = L_z = 30$)  ($t=1.0$, $m=-1.6$, $\delta=0.2$). 
The periodic and open boundary conditions are imposed along the $x$ and $z$ directions, respectively.
The infinitesimal number $\eta$ is chosen as $\eta = 1/\sqrt{L_z}$~\cite{eta}.}	
    \label{fig: 3dTI}
\end{figure}

\textit{Three dimensions}.---We further study a three-dimensional topological insulator described by~\cite{Schnyder-08}
\begin{align}
    \label{eq:3dTI}
    &H_{\mathrm{3DTI}} \left( \bm{k} \right) =
    \left( m + t \cos{k_x}+t \cos{k_y}+t \cos{k_z} \right) \tau_y  \nonumber \\
    &\qquad + \left( t \sin{k_x} \right) \sigma_x \tau_x
    + \left( t \sin{k_y} \right) \sigma_y \tau_x  \nonumber \\
    &\qquad + \left( t \sin{k_z} \right) \sigma_z \tau_x 
      + \delta \left( \cos{k_x} + \cos{k_y} \right) \sigma_y \tau_y
\end{align} 
with Pauli matrices $\sigma_i$'s and $\tau_i$'s ($i=x,y,z$), and real parameters $m, t, \delta \in \mathbb{R}$.
Similar to the SSH model, this Hamiltonian respects chiral symmetry $\tau_z H_{\mathrm{3DTI}} \left( \bm{k} \right) \tau_z = - H_{\mathrm{3DTI}} \left( \bm{k} \right)$ and is characterized by the three-dimensional winding number $W_3 \in \mathbb{Z}$~\cite{CTSR-review}.
In the topological phase, the nontrivial bulk topology $W_3 \neq 0$ leads to the emergence of Dirac surface states.
Applying the open boundary conditions along the $z$ direction, we regard the two-dimensional surface at $z=1$ as a system and the remaining bulk an environment (Fig.~\ref{fig: 3dTI}).
The effective surface Hamiltonian $H_{\rm eff} \left( E, k_x, k_y \right)$ hosts point-gap topology, characterized by the first Chern number of $\ii H_{\rm eff} \left( E, k_x, k_y \right) \tau_z$.
In contrast to the two-dimensional case, this point-gap topology is continuously deformable to (anti-)Hermitian topology, resulting in anomalous hinge states with the imaginary dispersion $\propto \ii k_x$ instead of the skin effect~\cite{OKSS-20, Nakamura-24}.
The extrinsic non-Hermitian topology also implies the qualitatively weaker bulk-boundary coupling than the intrinsic one for the two-dimensional case.

\begin{table}[t]
    \centering
     \caption{Periodic table of Hermitian topological insulators and superconductors in spatial dimensions $d=1, 2, 3$.
     The tenfold Altland-Zirnbauer (AZ) symmetry classes consist of time-reversal symmetry (TRS), particle-hole symmetry (PHS), and chiral symmetry (CS).
     In the entries specified by $^{*}$, effective non-Hermitian Hamiltonians at boundaries exhibit the skin effect.
     In 3D class DIII, specified by $^{**}$, the skin effect arises only for the odd number of the topological invariant~\cite{Nakamura-24}.}
    \begin{tabular}{c|ccc|ccc}
          \hline\hline
           ~AZ class~ & ~TRS~ & ~PHS~ & ~CS~ & ~$d=1$~ & ~$d=2$~ & ~$d=3$~ \\
      \hline 
        A & $0$ & $0$ & $0$ & $0$ & $\mathbb{Z}^{*}$ & $0$ \\
         AIII  & $0$ & $0$ & $1$ & $\mathbb{Z}$ & $0$ & $\mathbb{Z}$  \\
        \hline 
         AI & $+1$ & $0$ & $0$ & 0 &  0 & 0 \\
         BDI & $+1$ & $+1$ & $1$ & $\mathbb{Z}$ & 0 & 0 \\
         D & $0$ & $+1$ & $0$ & $\mathbb{Z}_2$ & $\mathbb{Z}^{*}$ & 0 \\
         DIII & $-1$ & $+1$ & $1$ & $\mathbb{Z}_2$ & $\mathbb{Z}_2^{*}$ & $\mathbb{Z}^{**}$ \\
         AII & $-1$ & $0$ & $0$ &  0 & $\mathbb{Z}_2^{*}$ & $\mathbb{Z}_2^{*}$ \\
         CII & $-1$ & $-1$ & $1$ & $2\mathbb{Z}$ & 0 & $\mathbb{Z}_2$   \\
         C & $0$ & $-1$ & $0$ & 0 &$2\mathbb{Z}^{*}$ & 0   \\
         CI & $+1$ & $-1$ & $1$ & 0 & 0 & $2\mathbb{Z}$  \\ \hline\hline
    \end{tabular}
    \label{tab:periodictable-h}
\end{table}

\textit{Classification}.---Anomalous boundary states in topological insulators generally host non-Hermitian topology, as summarized in Table~\ref{tab:periodictable-h}.
When the original Hermitian Hamiltonians in Eq.~(\ref{eq: Ham-tot}) respect the Altland-Zirnbauer (AZ) symmetry~\cite{AZ-97, CTSR-review}, the effective non-Hermitian Hamiltonians $H_{\rm eff} \left( E \right) = H_{\rm edge} + \Sigma \left( E \right)$ respect the corresponding symmetry known as the AZ$^{\dag}$ symmetry~\cite{supplement, KSUS-19}.
Consequently, $H_{\rm eff} \left( E \right)$ shows $d$-dimensional non-Hermitian topology inheriting from $\left( d+1 \right)$-dimensional Hermitian topology.
While symmetry of quadratic Lindbladians and entire Hermitian Hamiltonians was studied~\cite{Lieu-20}, we here uncover direct topological connection between closed and open quantum systems through the Green's function formalism, elucidating the microscopic physical origin underlying the mathematical classification~\cite{KSUS-19, JYLee-19, Bessho-21}.
In Table~\ref{tab:periodictable-h}, various boundary phenomena, as well as the distinct bulk-boundary coupling, are classified based on 
intrinsic and extrinsic non-Hermitian topology~\cite{OKSS-20, Nakamura-24}.
For example, helical Dirac surface states of time-reversal-invariant topological insulators host the $\mathbb{Z}_2$ skin effect, contrasting with the chiral-symmetric topological insulator in Eq.~(\ref{eq:3dTI}).

\textit{Discussions}.---In this Letter, we reveal inherent non-Hermitian topology of anomalous boundary states in Hermitian topological matter.
We demonstrate that the self-energy quantifying the particle exchange between the bulk and boundary detects Hermitian topology in the bulk and non-Hermitian topology at the boundary.
While point-gap topology of anomalous boundary states was investigated in the presence of non-Hermitian external perturbations~\cite{Ma-24, Schindler-23, Nakamura-23}, we here show non-Hermitian topology inherently embedded within Hermitian topological boundary states.

While this Letter focuses on clean noninteracting systems, our formalism should be extended to topological matter with disorder or many-body interactions, which we investigate in future work.
For example, we can revisit the interaction-induced reduction of topological classification from our non-Hermitian perspective~\cite{Fidkowski-10, *Fidkowski-11}.
Additionally, it merits further study to develop a field-theoretical understanding of our correspondence.
Our finding implies that anomaly of topological boundary states has close connection with a different type of anomaly accompanying non-Hermitian topology~\cite{KSR-21}.
It also provides a microscopic approach to realizing anomalous states and circumventing the fermion doubling theorem in non-Hermitian lattice models~\cite{NN-81a, *NN-81b, Karsten-81, Fujikawa-textbook, Armitage-review, Chen-23}.

\medskip
\begingroup
\renewcommand{\addcontentsline}[3]{}
\begin{acknowledgments}
We thank Akito Daido and Daichi Nakamura for helpful discussions.
S.H. thanks Hosho Katsura for letting him know Ref.~\cite{Sirker-14}.
K.K. thanks Taylor L. Hughes for helpful discussions and appreciates the program ``Topological Quantum Matter: Concepts and Realizations" held at Kavli Institute for Theoretical Physics (KITP), University of California Santa Barbara (National Science Foundation under Grant No.~NSF~PHY-1748958).
We appreciate the long-term workshops ``Quantum Information, Quantum Matter and Quantum Gravity" (YITP-T-23-01) and ``Recent Developments and Challenges in Topological Phases" (YITP-T-24-03) held at Yukawa Institute for Theoretical Physics (YITP), Kyoto University.
S.H. acknowledges travel support from MEXT KAKENHI Grant-in-Aid for Transformative Research Areas A ``Extreme Universe" collaboration (Grant Nos.~21H05182 and~22H05247).
S.H. is supported by JSPS Research Fellow No.~24KJ1445, JSPS Overseas Challenge Program for Young Researchers, and MEXT WISE Program.
S.H. and T.Y. are supported by JSPS KAKENHI Grant Nos.~21K13850 and 23KK0247.
S.H. and T.Y. are grateful for the support and hospitality of the Pauli Center for Theoretical Studies.
K.K. is supported by MEXT KAKENHI Grant-in-Aid for Transformative Research Areas A ``Extreme Universe" No.~24H00945.
\end{acknowledgments}
\endgroup

\let\oldaddcontentsline\addcontentsline
\renewcommand{\addcontentsline}[3]{}
\bibliography{ref.bib}

\begin{thebibliography}{137}
\bibitem [{eta()}]{eta}%
  \BibitemOpen
  \href@noop {} {}\bibinfo {note} {{$\eta = 1/\sqrt{L}$ is chosen to be much larger than the level spacing $\sim 1/L$ but much smaller than the macroscopic energy scale $\sim 1$.}}\BibitemShut {Stop}%
\end{thebibliography}
\let\addcontentsline\oldaddcontentsline

\clearpage
\widetext

\setcounter{secnumdepth}{3}

\renewcommand{\theequation}{S\arabic{equation}}
\renewcommand{\thefigure}{S\arabic{figure}}
\renewcommand{\thetable}{S\arabic{table}}
\setcounter{equation}{0}
\setcounter{figure}{0}
\setcounter{table}{0}
\setcounter{section}{0}
\setcounter{tocdepth}{0}

\numberwithin{equation}{section} 

\begin{center}
{\bf \large Supplemental Material for \\ \smallskip 
``Non-Hermitian Topology in Hermitian Topological Matter"}
\end{center}


\section{Effective non-Hermitian Hamiltonian}
    \label{asec: formalism}

We consider an edge Hamiltonian $H_{\rm edge}$ that is attached to a bulk Hamiltonian $H_{\rm bulk}$.
The whole system is described as
\begin{equation}
    H = \begin{pmatrix}
        H_{\rm bulk} & T \\
        T^{\dag} & H_{\rm edge}
    \end{pmatrix},
\end{equation}
where $T$ denotes the coupling between the bulk and the edge. 
The Green's function $G$ of $H$ is defined by~\cite{Datta-textbook}
\begin{equation}
    \left[ \left( E + \ii \eta \right) - H \right] G = 1,
\end{equation}
or
\begin{equation}
    \begin{pmatrix}
        E + \ii \eta - H_{\rm bulk} & -T \\
        -T^{\dag} & E + \ii \eta - H_{\rm edge}
    \end{pmatrix} \begin{pmatrix}
        G_{\rm bulk} & G_{\rm bulk-edge} \\
        G_{\rm edge-bulk} & G_{\rm edge}
    \end{pmatrix} = 1,
\end{equation}
with an infinitesimal number $\eta > 0$. 
This equation reads
\begin{align}
    \left( E + \ii \eta - H_{\rm bulk} \right) G_{\rm bulk-edge} - TG_{\rm edge} &= 0, \\
    -T^{\dag} G_{\rm bulk-edge} + \left( E + \ii \eta - H_{\rm edge} \right) G_{\rm edge} &= 1,
\end{align}
leading to
\begin{equation}
    G_{\rm bulk-edge} 
    = \left( E + \ii \eta - H_{\rm bulk} \right)^{-1} T G_{\rm edge}
    \eqqcolon G_{\rm bulk}^{(0)} T G_{\rm edge},
\end{equation}
with the Green's function $G_{\rm bulk}^{(0)} \coloneqq \left( E + \ii \eta - H_{\rm bulk} \right)^{-1}$ of the isolated bulk. 
Moreover, we have 
\begin{equation}
\left( E+ \ii \eta - H_{\rm edge} - \Sigma \right) G_{\rm edge} = 1,
\end{equation}
where we introduce the self-energy $\Sigma$ of the bulk as
\begin{equation}
\Sigma \coloneqq T^{\dag} \left( E + \ii \eta - H_{\rm bulk} \right)^{-1} T
= T^{\dag} G_{\rm bulk}^{(0)} T.
\end{equation}
The effective non-Hermitian Hamiltonian at the edge reads
\begin{equation}
    H_{\rm eff} \coloneqq H_{\rm edge} + \Sigma = H_{\rm edge} + T^{\dag} \left( E + \ii \eta - H_{\rm bulk} \right)^{-1} T.
\end{equation}
Let $E_{n}$ be an eigenenergy of $H_{\rm bulk}$ and $\ket{\psi_{n}}$ be the corresponding eigenstate. 
Then, we have the spectral decomposition
\begin{equation}
\Sigma = T^{\dag} \left( \sum_{n} \frac{\ket{\psi_{n}} \bra{\psi_{n}}}{E + \ii \eta - E_{n}}\right) T.
\end{equation}

\subsection*{Example}

As the simplest example, we consider a single-band model~\cite{Ryndyk-09}
\begin{equation}
    H_{ij} = t \left( \delta_{i, j+1} + \delta_{i, j-1} \right) \quad \left( t > 0 \right).
\end{equation}
Let $\ket{\psi_n} = ( \psi_{1}^{(n)}, \psi_{2}^{(n)}, \cdots, \psi_{L}^{(n)} )$ be an eigenstate of $H_{\rm bulk}$. 
Then, the Schr\"odinger equation $H \ket{\psi_n} = E_n \ket{\psi_n}$ reads
\begin{equation}
    t \psi_{j-1}^{(n)} + t\psi_{j+1}^{(n)} = E_n \psi_{j}^{(n)} \quad \left( j = 2, 3, \cdots, L-1 \right),\quad
    \psi_{0}^{(n)} = \psi_{L+1}^{(n)} = 0.
\end{equation}
The eigenenergies and eigenstates are given as
\begin{equation}
    E_{n} = 2t \cos k_{n},\quad
    \psi_{j}^{(n)} = \sqrt{\frac{2}{L+1}} \sin \left( k_{n} j \right);\quad
    k_{n} \coloneqq \frac{n\pi}{L+1}\quad\left( n = 1,2,\cdots, L \right).
\end{equation}
Thus, the self-energy of the bulk is obtained as
\begin{equation}
    \Sigma = t^{2} \sum_{n=1}^{L} \frac{|\psi_j^{(n)}|^2}{E+\ii \eta - E_{n}}
    = \frac{2t^{2}}{L+1} \sum_{n=1}^{L} \frac{\sin^{2} k_{n}}{E+\ii \eta - 2t\cos k_{n}}.
\end{equation}
In the semi-infinite limit $L \to \infty$, we have
\begin{equation}
\label{aeq:single-final}
\Sigma = 2t^{2} \sum_{n=1}^{L} \frac{\Delta k_{n}}{\pi} \frac{\sin^{2} k_{n}}{E+\ii \eta - 2t\cos k_{n}}
\to 2t^{2} \int_{0}^{\pi} \frac{dk}{\pi} \frac{\sin^{2} k}{E + \ii \eta - 2t \cos k}
= \frac{E}{2} \left( 1 - \sqrt{1 - \frac{4t^2}{E^2}}\right).
\end{equation}
The imaginary part of $\Sigma$ appears for $E$ within the bulk bandwidth $\left| E \right| < 2t$, which reflects the particle exchange between the bulk and the edge.

\section{Symmetry classification of non-Hermitian self-energy}
    \label{asec: symmetry}

We study symmetry of effective non-Hermitian Hamiltonians introduced in Sec.~\ref{asec: formalism}.
Specifically, we show that when original Hermitian Hamiltonians belong to the Altland-Zirnbauer symmetry class~\cite{AZ-97}, the corresponding effective non-Hermitian Hamiltonians belong to the Altland-Zirnbauer$^{\dag}$ symmetry class~\cite{KSUS-19}.
The similar correspondence was derived, for example, for quadratic Lindbladians~\cite{Lieu-20} and reflection matrices in scattering processes~\cite{Kawabata-23SVD}.
Because of this correspondence, the topological classification of $d$-dimensional Hermitian Hamiltonians coincides with that of $\left( d-1 \right)$-dimensional effective non-Hermitian Hamiltonians~\cite{KSUS-19, JYLee-19, Bessho-21, Schindler-23, Nakamura-23}.

\subsection{Time-reversal symmetry}

Suppose that the original Hermitian Hamiltonian $H$ respects time-reversal symmetry:
\begin{equation}
    \mathcal{T} H^{*} \mathcal{T}^{-1} = H
\end{equation}
with a unitary matrix $\mathcal{T}$.
Since time reversal acts only on the internal degrees of freedom, we have
\begin{equation}
    \mathcal{T} H^{*}_{\rm bulk} \mathcal{T}^{-1} = H_{\rm bulk},\quad
    \mathcal{T} H^{*}_{\rm edge} \mathcal{T}^{-1} = H_{\rm edge},\quad
    \mathcal{T} T^{*} \mathcal{T}^{-1} = T.
\end{equation}
Consequently, we have
\begin{equation}
    \mathcal{T} \Sigma^{T} \left( E \right) \mathcal{T}^{-1} = T^{\dag} \left( E + \ii \eta - H_{\rm bulk} \right)^{-1} T = \Sigma \left( E \right),
\end{equation}
and hence
\begin{equation}
    \mathcal{T} H_{\rm eff}^{T} \left( E \right) \mathcal{T}^{-1} = H_{\rm eff} \left( E \right).
\end{equation}
Thus, for arbitrary $E$, the effective non-Hermitian Hamiltonian $H_{\rm eff}$ respects time-reversal symmetry$^{\dag}$~\cite{KSUS-19}.
Notably, $H_{\rm eff}$ is not invariant under time reversal (i.e., $\mathcal{T} H_{\rm eff}^{*} \mathcal{T}^{-1} \neq H_{\rm eff}$).

\subsection{Particle-hole symmetry}

Suppose that the original Hermitian Hamiltonian $H$ respects particle-hole symmetry:
\begin{equation}
    \mathcal{C} H^{*} \mathcal{C}^{-1} = - H
\end{equation}
with a unitary matrix $\mathcal{C}$.
Since particle-hole transformation acts only on the internal degrees of freedom, we have
\begin{equation}
    \mathcal{C} H^{*}_{\rm bulk} \mathcal{C}^{-1} = - H_{\rm bulk},\quad
    \mathcal{C} H^{*}_{\rm edge} \mathcal{C}^{-1} = - H_{\rm edge},\quad
    \mathcal{C} T^{*} \mathcal{C}^{-1} = - T.
\end{equation}
Consequently, we have
\begin{equation}
    \mathcal{C} \Sigma^{*} \left( E \right) \mathcal{C}^{-1} = T^{\dag} \left( E - \ii \eta + H_{\rm bulk} \right)^{-1} T = - \Sigma \left( -E \right),
\end{equation}
and hence
\begin{equation}
    \mathcal{C} H_{\rm eff}^{*} \left( E \right) \mathcal{C}^{-1} = - H_{\rm eff} \left( -E \right).
\end{equation}
Thus, at the particle-hole-symmetric point $E=0$, the effective non-Hermitian Hamiltonian $H_{\rm eff} \left( E=0 \right)$ respects particle-hole symmetry$^{\dag}$~\cite{KSUS-19}.
Notably, $H_{\rm eff}$ does not respect particle-hole symmetry even at $E=0$ [i.e., $\mathcal{C} H_{\rm eff}^{T} \left( E=0 \right) \mathcal{C}^{-1} \neq - H_{\rm eff} \left( E = 0 \right)$].

\subsection{Chiral symmetry}

Suppose that the original Hermitian Hamiltonian $H$ respects chiral symmetry:
\begin{equation}
    \Gamma H \Gamma^{-1} = - H
\end{equation}
with a unitary matrix $\Gamma$.
Since chiral transformation acts only on the internal degrees of freedom, we have
\begin{equation}
    \Gamma H_{\rm bulk} \Gamma^{-1} = - H_{\rm bulk},\quad
    \Gamma H_{\rm edge} \Gamma^{-1} = - H_{\rm edge},\quad
    \Gamma T \Gamma^{-1} = - T.
\end{equation}
Consequently, we have
\begin{equation}
    \Gamma \Sigma^{\dag} \left( E \right) \Gamma^{-1} = T^{\dag} \left( E - \ii \eta + H_{\rm bulk} \right)^{-1} T =  - \Sigma \left( -E \right),
\end{equation}
and hence
\begin{equation}
    \Gamma H_{\rm eff}^{\dag} \left( E \right) \Gamma^{-1} = - H_{\rm eff} \left( -E \right).
\end{equation}
Thus, at the chiral-symmetric point $E=0$, the effective non-Hermitian Hamiltonian $H_{\rm eff} \left( E=0 \right)$ respects chiral symmetry~\cite{KSUS-19}.
Notably, $H_{\rm eff}$ does not respect sublattice symmetry even at $E=0$ [i.e., $\Gamma H_{\rm eff} \left( E=0 \right) \Gamma^{-1} \neq - H_{\rm eff} \left( E = 0 \right)$].

\section{Analytical derivation of the self-energy}
    \label{asec: analy-self}
    
We analytically derive the self-energy between the bulk and boundaries for the Su-Schrieffer-Heeger (SSH) model and a Chern insulator.

\subsection{Su-Schrieffer-Heeger model}

We consider the SSH model~\cite{SSH-79}
\begin{equation}
    H_{\mathrm{bulk}} \left( k \right) = \left( v + t \cos{k} \right) \sigma_x + \left( t \sin{k} \right) \sigma_y,
\end{equation}
where $\sigma_i$'s ($i=x, y, z$) are Pauli matrices, and $v>0$ and $t>0$ are the intracell and intercell hopping amplitudes, respectively.
Below, we assume the open boundary conditions.
In the topologically nontrivial phase $r<1$ ($r \coloneqq v/t$), a pair of zero-energy states appears at the boundaries.  
We assume that the total number of sites is odd, $2L+1$, for the sake of simplicity, for which eigenstates are exactly obtained under the open boundary conditions (see, for example, Refs.~\cite{Sirker-14, Shin-97}).
Let $\ket{\psi_n} = (\psi_1^{(n)}, \psi_2^{(n)}, \cdots, \psi_{2L+1}^{(n)})$ be an eigenstate.
The bulk states host the eigenenergy
\begin{equation}
    \label{e:ssh-bulk-energy}
  E_n = \pm t\sqrt{r^2+1+2r\cos{k_n}} \quad \left( k_n \coloneqq \frac{n\pi}{L+1};~n=1, \cdots, L\right),
\end{equation}
and the corresponding bulk eigenstates are obtained as
\begin{equation}
\label{eq:ssh-bulk-state}
  \psi_{2j-1}^{(n)} = \frac{r\sin \left( k_n j \right)+ \sin \left( k_n (j-1) \right)}{\sqrt{(L+1)
  \left( 1+2 r\cos{k_n}+r^2 \right)}} \hspace{5mm} \left(j=1,\cdots, L+1\right), \quad\quad ~\psi_{2j}^{(n)} = \pm \frac{\sin{\left( k_n j \right)}}{L+1} \hspace{5mm} \left(j=1,\cdots, L\right).
\end{equation}
On the other hand, the zero-energy eigenstate localized at the left edge is given as
\begin{equation}
\psi_{2j}=0 \hspace{5mm}\left(j=1,\cdots, L\right),\quad\quad \psi_{2j+1}= (-r)^j \sqrt{\frac{1-r^{2}}{1-r^{2(L+1)}}} \hspace{5mm}\left(j=0,\cdots, L\right)
\end{equation}
for $r<1$.

Now, we compute the self-energy $\Sigma = \Sigma_{\mathrm{edge}} + \Sigma_{\mathrm{bulk}}$. 
The contribution $\Sigma_{\mathrm{edge}}$ from the edge state is obtained as
\begin{equation}
    \label{aeq:ssh-edge-Linfty}
     \Sigma_{\mathrm{edge}} = \frac{t^2 (1-r^2) }{1-{r}^{2(L+1)}} \frac{1}{E+ \ii \eta} \frac{\sigma_0 - \sigma_z}{ 2} 
     \rightarrow 
       \frac{t^2 (1-r^2)}{E+ \ii \eta} \frac{\sigma_0 - \sigma_z}{2}
\end{equation}
in the limit $L \rightarrow \infty$ for $r < 1$.
Here, $\sigma_0$ is the $2 \times 2$ identity matrix.
Next, from Eqs.~\eqref{e:ssh-bulk-energy} and \eqref{eq:ssh-bulk-state}, 
the self-energy $\Sigma_{\mathrm{bulk}}$ 
from
the bulk states is obtained as
\begin{align}
\Sigma_{\mathrm{bulk}}=
    &\frac{t^2}{L+1}\sum\limits_{n=1}^{L}\frac{r^2\sin^2{k_n}}{r^2+1+2r\cos{k_n}}
    \Biggl(
    \frac{1}{E+\ii\eta-t\sqrt{r^2+1+2r\cos{k_n}}}
    +
     \frac{1}{E+\ii\eta+t\sqrt{r^2+1+2r\cos{k_n}}}
    \Biggr) \frac{\sigma_0 - \sigma_z}{2} \nonumber \\
    \rightarrow\,
    & t^2 
    \int_{0}^{\pi}\frac{dk}{\pi}
    \frac{r^2\sin^2{k}}{r^2+1+2r\cos{k}}
    \Biggl(
    \frac{1}{E+\ii\eta-t\sqrt{r^2+1+2r\cos{k}}}
    +
     \frac{1}{E+\ii\eta+t\sqrt{r^2+1+2r\cos{k}}}
    \Biggr)  \frac{\sigma_0 - \sigma_z}{2}
\end{align}
for $L \rightarrow \infty$.
By introducing the complex variable $z=e^{\ii k}$, the self-energy further leads to the contour integral
\begin{equation}
\label{aeq:ssh-bulk-Linfty}
    \Sigma_{\mathrm{bulk}} \rightarrow
    \frac{E+ \ii \eta}{2 } \frac{\sigma_0 - \sigma_z}{2}
     \frac{1}{2 \pi \ii} \oint_{C} dz~f(z),
\end{equation}
where $C$ is the unit circle in the complex plane, and $f(z)$ is defined as
\begin{equation}
    f(z) \coloneqq \frac{1}{z} \frac{(z^2-1)^2}{z^2+\left(r+r^{-1}\right)z+1}
    \frac{1}{z^2 + 2u z+1}, \quad 2u \coloneqq \left(r+\frac{1}{r}\right)-\frac{1}{r}\left(\frac{E+\ii\eta}{t}\right)^2.
\end{equation}
The poles of $f(z)$ are 
\begin{equation}
    z = 0, - r, -\frac{1}{r}, z_{\pm},\quad \left( z_{\pm} \coloneqq -u \pm \sqrt{u^2-1} \right),
\end{equation}
whose residues are calculated as
\begin{equation}
     \Res [z=0] = 1,~ \Res[z=-r] = -\frac{t^2 (1- r^2) }{\left({E+\ii \eta}\right)^2},
     ~\Res[z=-\frac{1}{r}] = \frac{t^2 (1 - r^2) }{\left({E+\ii \eta}\right)^2},
     ~ \Res [z=z_\pm] = \pm \frac{2 t^2 r \sqrt{u^2-1}}{\left({E+\ii \eta}\right)^2}.
\end{equation}
Depending on $r$, each pole is either located inside or outside the unit circle $C$,
as summarized in the following:
\begin{enumerate}[(i)]
    \item For $r > 1$
    and $E^2 < t^2 (r^2 + 1)$, 
    only the three poles $z=0, -r^{-1}, z_+$ are located inside the unit circle $C$, 
    leading to
    \begin{equation}
         \Sigma_{\mathrm{bulk}} =  \left(
    \frac{E+\ii\eta}{2} +\frac{1}{2} \frac{t^2 (1-r^2)}{E+\ii\eta}
    + \frac{1}{2} \frac{\sqrt{\left[ t^2(r^2+1)-\left(E+\ii\eta\right)^2 \right]^2-4 t^4 r^2}}{E+\ii\eta} \right) \frac{\sigma_0 - \sigma_z}{2}.
    \end{equation}
    
    \item For $r > 1$
    and $E^2 > t^2 (r^2 + 1)$, 
    only the three poles $z=0, -r^{-1}, z_-$ are located inside the unit circle $C$, 
    leading to
    \begin{equation}
         \Sigma_{\mathrm{bulk}} =  \left(
    \frac{E+\ii\eta}{2} +\frac{1}{2} \frac{t^2 (1-r^2)}{E+\ii\eta}
    - \frac{1}{2} \frac{\sqrt{\left[ t^2(r^2+1)-\left(E+\ii\eta\right)^2 \right]^2-4 t^4 r^2}}{E+\ii\eta} \right) \frac{\sigma_0 - \sigma_z}{2}.
    \end{equation}
    
    \item For $r < 1$
    and $E^2 < t^2 (r^2 + 1)$, 
    only the three poles $z=0, -r, z_+$ are located inside the unit circle $C$, 
    leading to
    \begin{equation}
         \Sigma_{\mathrm{bulk}} =  \left(
    \frac{E+\ii\eta}{2} -\frac{1}{2} \frac{t^2 (1-r^2)}{E+\ii\eta}
    + \frac{1}{2} \frac{\sqrt{\left[ t^2(r^2+1)-\left(E+\ii\eta\right)^2 \right]^2-4 t^4 r^2}}{E+\ii\eta} \right) \frac{\sigma_0 - \sigma_z}{2}.
    \end{equation}
    
    \item For $r < 1$
    and $E^2 > t^2 (r^2 + 1)$, 
    only the three poles 
    $z=0, -r, z_-$ are located inside the unit circle $C$, 
    leading to
    \begin{equation}
         \Sigma_{\mathrm{bulk}} = \left(
    \frac{E+\ii\eta}{2} -\frac{1}{2} \frac{t^2 (1-r^2)}{E+\ii\eta}
    - \frac{1}{2} \frac{\sqrt{\left[ t^2(r^2+1)-\left(E+\ii\eta\right)^2 \right]^2-4 t^4 r^2}}{E+\ii\eta} \right) \frac{\sigma_0 - \sigma_z}{2}.
    \end{equation}
\end{enumerate}

Combining both contributions from the bulk and edge, we obtain the total self-energy as
\begin{equation}
\label{aeq:ssh-self-final}
    \Sigma (E) = 
    \left(
    \frac{E+\ii\eta}{2} +\frac{1}{2} \frac{t^2 (1-r^2)}{E+\ii\eta}
    +\frac{1}{2}\mathrm{sgn} \left[t^2 (r^2 + 1) -E^2 \right] \frac{\sqrt{\left[ t^2(r^2+1)-\left(E+\ii\eta\right)^2 \right]^2-4 t^4 r^2}}{E+\ii\eta} \right) \frac{\sigma_0 - \sigma_z}{2}.
\end{equation}
The imaginary part of the self-energy appears for $E$ within the bulk bandwidth $\left|v-t\right| <  \left| E\right| < \left|v+t\right|$
or for zero energy $E=0$.
At the critical point $r=1$,
Eq.~\eqref{aeq:ssh-self-final} reduces to Eq.~\eqref{aeq:single-final}.
Additionally, at zero energy $E=0$, the self-energy reduces to
\begin{equation}
    \Sigma (0) =
    \begin{cases}
        0 \hspace{5mm} & (r>1); \\
        -\ii \pi t^2 (1-r^2) \delta(0) \left( \sigma_0 - \sigma_z \right)/2 \hspace{5mm} & (r<1).
    \end{cases}
\end{equation}
Thus, in the topologically nontrivial phase ($r<1$), the imaginary part of the self-energy diverges.

\subsection{Chern insulator}
We consider the Chern insulator~\cite{QWZ-06}
\begin{equation}
\label{aeq:qwz}
    H_{\mathrm{bulk}} \left( k_x, k_y \right) = \left( t \sin{k_x} \right) \sigma_x + \left( t \sin{k_y} \right) \sigma_y + \left( m + t \cos{k_x}+t \cos{k_y} \right) \sigma_z
\end{equation}
with Pauli matrices $\sigma_i$'s ($i=x,y,z$).
Here, $t > 0$, denotes the hopping amplitude, and $m \in \mathbb{R}$ denotes the onsite potential.
We assume the square lattice under the open boundary conditions along the $x$ direction and the periodic boundary conditions along the $y$ direction.
Defining $R(k_y)$ and $S$ as
\begin{equation}
    R(k_y) \coloneqq \left( t \sin{k_y} \right) \sigma_y + \left( m+t \cos{k_y} \right) \sigma_z, \quad  S \coloneqq \frac{t}{2} \left( \sigma_z-\ii \sigma_x \right)
        \label{eq:RS},
\end{equation}
we express the Hamiltonian as
\begin{equation}
    H_{\mathrm{bulk}}(k_y) = \sum\limits_{x=1}^{L} \ket{x}\bra{x} \otimes R(k_y) + 
     \sum\limits_{x=1}^{L-1}\Bigg[ \ket{x}\bra{x+1} \otimes S + 
     \ket{x+1}\bra{x} \otimes S^{\dagger}
     \Bigg].
\end{equation}
In the topologically nontrivial phase $\abs{m}<2t$, a chiral edge state appears around the boundaries $x = 1$ and $x=L$.
We use the ansatz
\begin{equation}
    \ket{\psi_{\mathrm{edge}}\left( k_y \right)}=\sum\limits_{x=1}^{L} \beta^{x} \left( k_y \right) \ket{x} \otimes \ket{u \left( k_y \right)}
\end{equation}
for the chiral edge state localized around $x=1$ with a 
$k_y$-dependent
two-component vector $\ket{u \left( k_y \right)}$.
Here, $-1/\log \left| \beta \left( k_y \right) \right|$ gives a $k_y$-dependent localization length of the chiral edge state along the $x$ direction.
Then, we reduce the eigenvalue equation $H_{\mathrm{bulk}}(k_y)\ket{\psi_{\mathrm{edge}}(k_y)} = \lambda_{\mathrm{edge}}(k_y)\ket{\psi_{\mathrm{edge}}(k_y)}$ to
\begin{equation}
    \left( R \left( k_y \right) +S \beta \left( k_y \right) + S^{\dag} \beta^{-1} \left( k_y \right) \right) \ket{u \left( k_y \right)} =  \lambda_{\mathrm{edge}}(k_y) \ket{u \left( k_y \right)} \hspace{10mm} (2 \leq x \leq L-1) \label{eq:eigen-bulk-orbital} 
\end{equation}
in the bulk and 
\begin{equation}
     \left( R \left( k_y \right) + S \beta \left( k_y \right) \right) \ket{u \left( k_y \right)} =  \lambda_{\mathrm{edge}}(k_y) \ket{u \left( k_y \right)} \hspace{10mm} (x = 1) \label{eq:eigen-edge-orbital}
\end{equation}
at the left edge. 
From these equations, we have
\begin{equation}
\label{eq:edge-henkei}
    S^{\dagger}\ket{u \left( k_y \right)} = 0,
\end{equation}
further leading to
\begin{equation}
\label{eq:u}
    \ket{u \left( k_y \right)} = \left(
    \begin{array}{c}
         1 \\
         \ii 
    \end{array}
    \right).
\end{equation}
Combining Eqs.~\eqref{eq:eigen-edge-orbital} and~\eqref{eq:u}, we obtain
\begin{equation}
\label{eq:eigen-totyu-1}
    \left[ - \frac{\ii t}{2}\beta \left( k_y \right) \sigma_x + \left( t \sin k_y \right)
\sigma_y+\left( m+\cos k_y + \frac{t}{2} \beta \left( k_y \right) \right) \sigma_z \right] \ket{u} =  \lambda_{\mathrm{edge}}(k_y) \ket{u}.
\end{equation}
From Eq.~\eqref{eq:eigen-totyu-1}, 
$\lambda_{\mathrm{edge}} \left( k_y \right)$ and $\beta \left( k_y \right)$ are given as
\begin{equation}
    \lambda_{\mathrm{edge}}(k_y) = t\sin{k_y}, \quad \beta(k_y) = -\left(\frac{m}{t} + \cos k_y\right).
\end{equation}
Thus, the self-energy is calculated as 
\begin{equation}
\label{aeq:qwz-final}
    \Sigma_{\mathrm{edge}} (k_y) = \frac{t^2(1-\beta^2)}{2(E+ \ii \eta-t\sin k_y)} (\sigma_0 -\sigma_y).
\end{equation}
For $k_y = 0$ or $k_y = \pi$, this self-energy essentially reduces to that of the SSH model [see Eq.~(\ref{aeq:ssh-edge-Linfty})].
Notably, these expressions are valid only for
\begin{equation}
\label{eq:inequality}
    1 > \abs{\beta} = \abs{\frac{m}{t}+\cos {k_y}}
\end{equation}
so that the above chiral edge state can be normalized.

\section{Current formula}

We derive the current formula in Eq.~\eqref{eq:current-formula} of the main text (see also 
``Chapter~8: Non-equilibrium Green's function formalism'' in Ref.~\cite{Datta-textbook}, as well as
``Chapter~9: Coherent transport" and ``Appendix: advanced formalism'' in Ref.~\cite{Datta-textbook-2005}).
We consider a noninteracting fermionic Hamiltonian described by
\begin{equation}
    \hat{H} = \sum_{i,j} \hat{a}_i ^\dag  H_{ij} \hat{a}_j ,
\end{equation}
where $\hat{a}_i$ ($\hat{a}^\dag_i$) is a fermionic annihilation (creation) operator at site $i$,
and $H_{ij}$ is a single-particle
Hermitian 
Hamiltonian. 
The Heisenberg equation for the annihilation operator
reads
\begin{equation}
    \ii \hbar \frac{d \hat{a}_i (t)}{dt} = [\hat{a}_i (t), \hat{H}] = \sum_j H_{ij}  \hat{a}_j (t).
\end{equation}
We rewrite this equation as
\begin{equation}
\label{aeq:heisenberg-total}
    \ii \hbar \frac{d}{dt}
    \begin{pmatrix}
   \hat{\boldsymbol{a}}_{\mathrm{bulk}}(t) \\
    \hat{\boldsymbol{a}}_{\mathrm{edge}}(t)
\end{pmatrix}
    = \begin{pmatrix}
        H_{\rm bulk} & T \\
        T^{\dag} & H_{\rm edge}
    \end{pmatrix}
    \begin{pmatrix}
   \hat{\boldsymbol{a}}_{\mathrm{bulk}}(t) \\
    \hat{\boldsymbol{a}}_{\mathrm{edge}}(t)
\end{pmatrix},
\end{equation}
where $\hat{\boldsymbol{a}}_{\mathrm{bulk}}(t)$ [$\hat{\boldsymbol{a}}_{\mathrm{edge}}(t)$] is a row vector
of annihilation operators 
acting on the bulk (edge).
We expand the annihilation operator for the bulk as
\begin{equation}
    \hat{\boldsymbol{a}}_{\mathrm{bulk}}(t) = \hat{\boldsymbol{a}}_{\mathrm{bulk}}^{(0)}(t) + \hat{\boldsymbol{\chi}}(t),
\end{equation}
where $\hat{\boldsymbol{a}}_{\mathrm{bulk}}^{(0)}(t)$ denotes the unperturbed annihilation operator satisfying $\ii \hbar \left( d/dt \right) \hat{\boldsymbol{a}}_{\mathrm{bulk}}^{(0)}(t) = H_{\rm bulk} \hat{\boldsymbol{a}}_{\mathrm{bulk}}^{(0)}(t)$.
From the original Heisenberg equation in Eq.~\eqref{aeq:heisenberg-total}, we have 
\begin{align}
    \left( \ii \hbar \frac{d}{dt} - H_{\rm bulk} \right) \hat{\boldsymbol{\chi}}(t) &= T \hat{\boldsymbol{a}}_{\mathrm{edge}}(t), 
        \label{aeq:Heisenberg-bulk} \\
    \left( \ii \hbar \frac{d}{dt} - H_{\rm edge} \right)\hat{\boldsymbol{a}}_{\mathrm{edge}}(t) &= T^\dag \left( \hat{\boldsymbol{a}}_{\mathrm{bulk}}^{(0)}(t) + \hat{\boldsymbol{\chi}}(t) \right).
        \label{aeq:Heisenberg-edge}
\end{align}
We introduce the retarded Green's function $g^{(0)}_{\mathrm{bulk}}(t-t')$ for the bulk 
\begin{equation}
\label{aeq:green-bulk}
     \left( \ii \hbar \frac{d}{dt} - H_{\rm bulk} \right) g^{(0)}_{\mathrm{bulk}}(t-t')
     =
    \delta (t-t').
\end{equation}
Here, the argument of the Green's function depends only on the time difference $t-t'$ since the Hamiltonian $H_{\mathrm{bulk}}$ is independent of time.
Using the Green's function, we obtain the formal solution of Eq.~\eqref{aeq:Heisenberg-bulk} by the convolution integral,
\begin{equation}
\label{aeq:chi-sol}
    \hat{\boldsymbol{\chi}}(t) = \int dt' g^{(0)}_{\mathrm{bulk}}(t-t')
    T \hat{\boldsymbol{a}}_{\mathrm{edge}}(t'),
\end{equation}
and hence
\begin{equation}
\label{aeq:Heisenberg-edge2}
     \left( \ii \hbar \frac{d}{dt} - H_{\rm edge} \right) \hat{\boldsymbol{a}}_{\mathrm{edge}}(t) 
     =
     T^\dag  \hat{\boldsymbol{a}}_{\mathrm{bulk}}^{(0)}(t) 
     + \int dt' \Sigma(t-t') \hat{\boldsymbol{a}}_{\mathrm{edge}}(t')
\end{equation}
with the self-energy defined as
\begin{equation}
    \Sigma(t-t') \coloneqq T^\dag g^{(0)}_{\mathrm{bulk}}(t-t') T.
\end{equation}
Similarly, introducing the retarded Green's function $g_{\mathrm{edge}}(t-t')$ for the edge
by
\begin{equation}
\label{aeq:green-edge}
     \left( \ii \hbar \frac{d}{dt} - H_{\rm edge} \right) g_{\mathrm{edge}}(t-t')  
     =
     \int dt'' \Sigma(t-t'') g_{\mathrm{edge}}(t''-t') 
     + \delta(t-t'),
\end{equation}
we obtain the formal solution of Eq.~\eqref{aeq:Heisenberg-edge2} as
\begin{equation}
\label{aeq:annhilation-sol}
    \hat{\boldsymbol{a}}_{\mathrm{edge}}(t)
    = \int dt' g_{\mathrm{edge}}(t-t') T^\dag  \hat{\boldsymbol{a}}_{\mathrm{bulk}}^{(0)}(t').
\end{equation}
Similar calculations for the creation operator lead to
\begin{equation}
\label{aeq:Heisenberg-edge-creation}
     \left( \ii \hbar \frac{d}{dt} + H_{\rm edge}^T \right) \hat{\boldsymbol{a}}_{\mathrm{edge}}^\dag(t) 
     =
     -\int dt' \Sigma^* (t-t') \hat{\boldsymbol{a}}_{\mathrm{edge}}^\dag(t')
     -
     T^T  \hat{\boldsymbol{a}}_{\mathrm{bulk}}^{(0)^{\dag}} (t)
\end{equation}
and hence
\begin{equation}
\label{aeq:creation-sol}
     \hat{\boldsymbol{a}}_{\mathrm{edge}}^\dag(t)
    = \int dt' g_{\mathrm{edge}}^* (t-t') T^T  \hat{\boldsymbol{a}}_{\mathrm{bulk}}^{(0)^{\dag}}(t').
\end{equation}
We define the Fourier transform of $f(t-t')$ as
\begin{equation}
    \mathcal{F}\left[f(t-t') \right] \coloneqq \int d(t-t') e^{\ii \frac{E}{\hbar}(t-t')} f(t-t').
\end{equation}
The Fourier transforms of the Green's functions $g^{(0)}_{\mathrm{bulk}}(t-t')$ in Eq.~(\ref{aeq:green-bulk}) and $g_{\mathrm{edge}}(t-t')$ in Eq.~(\ref{aeq:green-edge}) are
\begin{equation}
    \label{aeq: Green-bulk-zero}
    G_{\mathrm{bulk}}^{(0)}(E) \coloneqq \mathcal{F}\left[g_{\mathrm{bulk}}^{(0)}(t-t')\right] = \frac{1}{E + \ii \eta -H_{\mathrm{bulk}}}, \quad
    G_{\mathrm{edge}}(E) \coloneqq \mathcal{F}\left[g_{\mathrm{edge}}(t-t') \right] = \frac{1}{E+\ii \eta -H_{\mathrm{edge}}-\Sigma (E)}, 
\end{equation}
with $\Sigma(E) \coloneqq \mathcal{F}\left[\Sigma(t-t') \right]$.
Here, an infinitesimal number $\eta > 0$ is introduced so that the initial condition can be satisfied appropriately.

Now, we introduce the two-time current operator~\cite{Datta-textbook,Datta-textbook-2005}
\begin{equation}
    J(t, t') \coloneqq \left( \frac{d}{dt} + \frac{d}{dt'} \right) C(t,t'),
\end{equation}
with the two-time correlation function 
\begin{equation}
    C_{ij}(t,t') \coloneqq \langle \hat{a}_j^\dag (t') \hat{a}_i (t) \rangle,
\end{equation} 
where the bracket denotes the average over a thermal equilibrium state.
In the limit $t' \rightarrow t$, the diagonal element of $J (t,t')$ gives the local current.
From Eqs.~\eqref{aeq:annhilation-sol} and~\eqref{aeq:Heisenberg-edge-creation}, the correlation function of the edge, 
$C_{\mathrm{edge}; ij}(t,t') \coloneqq  \langle \hat{a}_{{\mathrm{edge}};j}^\dag (t') \hat{a}_{{\mathrm{edge}};i} (t) \rangle$, 
satisfies
\begin{equation}
\label{aeq:edge-correlation-diff1}
    \ii \hbar \frac{d}{dt} C_{\mathrm{edge}}(t,t') = 
    - C_{\mathrm{edge}}(t,t') H_{\mathrm{edge}}
    -
    \int dt'' g_{\mathrm{edge}} (t',t'') T^\dag C_{\mathrm{bulk}}^{(0)}(t, t'') T 
    -
    \int dt'' C_{\mathrm{edge}}(t'',t') \Sigma^\dag (t,t'') ,
\end{equation}
with the isolated bulk correlation function 
$C_{\mathrm{bulk}; ij}^{(0)}(t, t') \coloneqq \langle  \hat{a}^{(0)^{\dag}}_{\mathrm{bulk};j}(t') \hat{a}^{(0)}_{\mathrm{bulk};i} (t) \rangle$.
Here, we assume that the bulk maintains the local equilibrium even after attaching the edge~\cite{Datta-textbook, Datta-textbook-2005}.
Similarly, from Eqs.~\eqref{aeq:Heisenberg-edge2} and \eqref{aeq:creation-sol}, we have
\begin{equation}
\label{aeq:edge-correlation-diff2}
    \ii \hbar \frac{d}{dt'} C_{\mathrm{edge}}(t,t') = H_{\mathrm{edge}}
    C_{\mathrm{edge}}(t,t')  
    +
    \int dt''  T^\dag C_{\mathrm{bulk}}^{(0)}(t'', t')  T g_{\mathrm{edge}}^\dag (t,t'')
    +
    \int dt''  \Sigma (t',t'') C_{\mathrm{edge}}(t,t'') . 
\end{equation}
The first term of the right-hand side in Eqs.~\eqref{aeq:edge-correlation-diff1} and~\eqref{aeq:edge-correlation-diff2} represents the particle current inside the edge; we below omit it since it is irrelevant to the current between the bulk and edge, and hence non-Hermitian topology.
Additionally, the second (third) term in Eqs.~\eqref{aeq:edge-correlation-diff1} and~\eqref{aeq:edge-correlation-diff2} describes the particle current from the bulk to the edge (from the edge to the bulk).
Then, the net current is 
\begin{equation}
    J(t,t') = J_{\mathrm{inflow}} (t,t') - J_{\mathrm{outflow}} (t,t')
\end{equation}
with
\begin{align}
    \ii \hbar J_{\mathrm{inflow}} (t,t') 
    &\coloneqq 
    -\int dt'' g_{\mathrm{edge}} (t',t'') T^\dag C_{\mathrm{bulk}}^{(0)}(t, t'') T 
    +\int dt'' 
    T^\dag C_{\mathrm{bulk}}^{(0)}(t'', t')  T g_{\mathrm{edge}}^\dag (t,t''), \\
    \ii \hbar J_{\mathrm{outflow}} (t,t') 
    &\coloneqq 
        \int dt''  C_{\mathrm{edge}}(t'',t')   \Sigma^\dag (t,t'')
    -    \int dt'' \Sigma (t',t'') C_{\mathrm{edge}}(t,t'').
\end{align}
At thermal equilibrium, from the cyclicity of the trace, the correlation functions $C_{\mathrm{bulk}}^{(0)}(t, t')$ and $C_{\mathrm{edge}}(t,t')$ depend only on the argument $t-t'$.
After the Fourier transformation, the $E$-resolved currents read,
\begin{align}
    \ii \hbar J_{\mathrm{inflow}}(E) &\coloneqq \mathcal{F}\left[J_{\mathrm{inflow}}(t-t')\right] =  - G_{\mathrm{edge}}(E) T^\dag C_{\mathrm{bulk}}^{(0)} (E)  T 
    +  T^\dag  C_{\mathrm{bulk}}^{(0)} (E) T G_{\mathrm{edge}}^\dag (E), \\
    \ii \hbar J_{\mathrm{outflow}}(E) &\coloneqq \mathcal{F} \left[ J_{\mathrm{outflow}}(t-t') \right] = 
      C_{\mathrm{edge}} (E) \Sigma^\dag (E) -  \Sigma (E) C_{\mathrm{edge}} (E).
\end{align}

To calculate the correlation function $C_{\mathrm{bulk}}^{(0)} (E)$ of the isolated bulk, 
we diagonalize the Hamiltonian, 
\begin{equation}
    \hat{H}_{\mathrm{bulk}} = \sum_n E_n \hat{\gamma}^{{(0)}^\dag}_{\mathrm{bulk};n} \hat{\gamma}^{(0)}_{\mathrm{bulk};n}, \quad \hat{\gamma}^{(0)^\dag}_{\mathrm{bulk};n} \coloneqq \sum_m \hat{a}^{(0)^\dag}_{\mathrm{bulk};m} U_{mn},
\end{equation}
where $E_n$ is eigenenergy of the single-particle Hamiltonian $H_{\rm bulk}$, and $U_{mn}$ is a unitary matrix of the single-particle eigenstates.
Then, we have
\begin{equation}
    C_{\mathrm{bulk};ij}^{(0)} (E) = 
    \sum\limits_k U_{ik} U^\dag _{kj} \delta(E-E_k) \langle \hat{\gamma}^{(0)^\dag}_{\mathrm{bulk};k} \hat{\gamma}^{(0)}_{\mathrm{bulk};k} \rangle = f_0 \left( E \right) \sum_{k} U_{ik} U_{kj}^{\dag} \delta \left( E-E_k \right),
\end{equation}
with the Fermi distribution function $f_0 \left( E_k \right) = \langle \hat{\gamma}^{(0)^\dag}_{\mathrm{bulk};k} \hat{\gamma}^{(0)}_{\mathrm{bulk};k} \rangle$.
Using the Green's function in Eq.~(\ref{aeq: Green-bulk-zero}), i.e.,
\begin{equation}
    G_{\mathrm{bulk}; ij}^{(0)}(E) = \sum_{k} \frac{U_{ik} U^{\dag}_{kj}}{E - E_{k} + \ii \eta} = \sum_{k} U_{ik} U^{\dag}_{kj} \left[ \mathcal{P} \left( \frac{1}{E - E_{k}} \right) - \ii \pi \delta \left( E - E_{k} \right) \right],
\end{equation}
we have
\begin{equation}
    C_{\mathrm{bulk}}^{(0)} (E) = \frac{\ii f_0 (E)}{2\pi} \left[ G_{\mathrm{bulk}}^{(0)}(E) - G_{\mathrm{bulk}}^{(0)^\dag}(E) \right],
\end{equation}
and hence
\begin{equation}
    \label{aeq:TdagCT}
    T^\dag C_{\mathrm{bulk}}^{(0)} (E) T = \frac{\ii f_0 (E)}{2\pi} \left[ \Sigma(E) - \Sigma^\dag (E) \right] = \frac{f_0 (E)}{2\pi} \Gamma(E),
\end{equation}
with the broadening matrix
\begin{equation}
    \Gamma (E) \coloneqq  \ii \left[ \Sigma(E) - \Sigma^\dag (E) \right].
\end{equation}

Next, from Eqs.~\eqref{aeq:annhilation-sol} and~\eqref{aeq:creation-sol}, we have
\begin{equation}
    C_{\mathrm{edge}} (t,t') 
    = \int dt_1 \int dt_2 
    g_{\mathrm{edge}} (t',t_2)
    T^\dag 
    C_{\mathrm{bulk}}^{(0)} (t_2,t_1) T
    g_{\mathrm{edge}}^\dag (t,t_1)
\end{equation}
and hence
\begin{equation}
    C_{\mathrm{edge}} (E)
    = G_{\mathrm{edge}} (E) T^\dag C_{\mathrm{bulk}}^{(0)} (E) T G_{\mathrm{edge}}^\dag (E),
\end{equation}
further leading to
\begin{equation}
    C_{\mathrm{edge}} (E)
    =  \frac{f_0 (E)}{2\pi} G_{\mathrm{edge}} (E) \Gamma (E) G_{\mathrm{edge}}^\dag (E) = \frac{f_0 (E)}{2\pi}  A_{\mathrm{edge}}(E), 
        \label{aeq:edge-correlatuion-sol}
\end{equation}
where $A_{\mathrm{edge}}(E)$ is the spectral function 
\begin{equation}
    A_{\mathrm{edge}}(E) \coloneqq \ii \left[G_{\mathrm{edge}} (E) - G_{\mathrm{edge}}^\dag (E)\right] = G_{\mathrm{edge}} (E) \Gamma(E) G_{\mathrm{edge}}^\dag (E).
\end{equation}
Then, from Eqs.~\eqref{aeq:TdagCT} and~\eqref{aeq:edge-correlatuion-sol}, we obtain
\begin{align}
    J_{\mathrm{inflow}} (E) &= \frac{f_0 (E)}{2\pi\ii \hbar}
    \left[
    \Gamma(E) G_{\mathrm{edge}}^\dag (E) - G_{\mathrm{edge}} (E) \Gamma(E)
    \right], \\
    J_{\mathrm{outflow}} (E) &= \frac{f_0 (E)}{2\pi\ii \hbar} 
    \left[
     A_{\mathrm{edge}} (E) \Sigma^\dag (E) -\Sigma(E) A_{\mathrm{edge}}(E)
    \right],
\end{align}
and hence 
\begin{equation}
\label{aeq:E-resolved-current}
    J(E) = 
    J_{\mathrm{inflow}} (E) - J_{\mathrm{outflow}} (E) = 
    \frac{f_0 (E)}{2\pi \hbar} \left(
    \left[ \Sigma(E),G_{\mathrm{edge}}(E)\right] 
    +
    \left[ \Sigma(E),G_{\mathrm{edge}}(E)\right]^\dag
    \right)
\end{equation}
with the effective non-Hermitian Hamiltonian $H_{\mathrm{eff}} (E) \coloneqq H_{\mathrm{edge}}  + \Sigma(E)$. 
From the relation 
\begin{equation}
\label{aeq:current-final-mae}
    \big[
   \Sigma, G_{\mathrm{edge}} 
   \big]
   +
   \big[
   \Sigma, G_{\mathrm{edge}} 
   \big]^\dag
   = 
   - \big[
   H_{\mathrm{edge}}, G_{\mathrm{edge}} 
   \big]
   - \big[
   H_{\mathrm{edge}}, G_{\mathrm{edge}} 
   \big]^\dag,
\end{equation}
Eq.~\eqref{aeq:E-resolved-current} reduces to Eq.~\eqref{eq:current-formula} in the main text. 
In passing, ``$f_0/2 \pi \hbar$" is dropped in Eq.~\eqref{eq:current-formula}.

We also introduce the $E$-resolved current density $j(y,E)$ at site $y$, 
summing up the internal degree of freedom $\alpha$ as
\begin{equation}
\label{aeq:local-E-resolved-current}
    j(y,E) \coloneqq
     \sum\limits_{\alpha}
  \bra{y,\alpha}
     \left(
   \big[
   \Sigma, G_{\mathrm{edge}} 
   \big]
   +
   \big[
   \Sigma, G_{\mathrm{edge}} 
   \big]^\dag
   \right)
    \ket{y,\alpha}.
\end{equation}
Notably, while $j \left( y,E \right)$ can be nonzero, the net current $\sum_{y=1}^L j \left( y,E \right)$ flowing between the bulk and edge always vanishes owing to the cyclicity of the trace:
\begin{equation}
    \sum_{y=1}^L j \left( y,E \right) = 
    \Tr \left( - \big[
   H_{\mathrm{edge}}, G_{\mathrm{edge}} 
   \big]
   - \big[
   H_{\mathrm{edge}}, G_{\mathrm{edge}} 
   \big]^\dag \right) = 0,
\end{equation}
where the trace is taken over both sites $y$ and the internal degrees $\alpha$ of freedom.
This is consistent with the absence of net current at equilibrium~\cite{Datta-textbook-2005}.

Additionally, the current density $j(y,E)$ vanishes under the periodic boundary conditions along the $y$ direction.
Owing to translation invariance, the Hamiltonians can be written as
\begin{equation}
    H_{\mathrm{edge}} = \sum_{k; \alpha, \beta} h_{\mathrm{edge}}^{\alpha \beta} (k) \ket{k,\alpha} \bra{k, \beta},\quad  H_{\mathrm{eff}} = \sum_{k; \alpha, \beta} h_{\mathrm{eff}}^{\alpha \beta} (k) \ket{k,\alpha} \bra{k, \beta},
\end{equation}
where $k$ is momentum, and $\alpha$ and $\beta$ specify the internal degree of freedom.
Then, from the cyclicity of the trace, we have
\begin{equation}
\label{aeq:current-commute}
  \Trace_\alpha [h_{\mathrm{edge}}(k),h_{\mathrm{eff}}(k)] = 0,
\end{equation}
where the trace is taken only for the internal degree of freedom. 
Hence, we have $\Trace_\alpha [H_{\mathrm{edge}},H_{\mathrm{eff}}] = 0$ and
\begin{equation}
    j (y, E) = - \bra{y} \Trace_\alpha \left( [H_{\mathrm{edge}},G_{\mathrm{edge}}] + [H_{\mathrm{edge}},G_{\mathrm{edge}}]^{\dag} \right) \ket{y} = 0.
\end{equation}
This is consistent with the absence of the local current under the periodic boundary conditions, as shown in Fig.~\ref{fig:current}\,(c) of the main text.

\section{$\mathbb{Z}_2$ skin effect in time-reversal-invariant topological insulators}
    \label{asec: z2-skin}
    
\begin{figure}[hbt]
\centering
\includegraphics[width=0.3\linewidth]{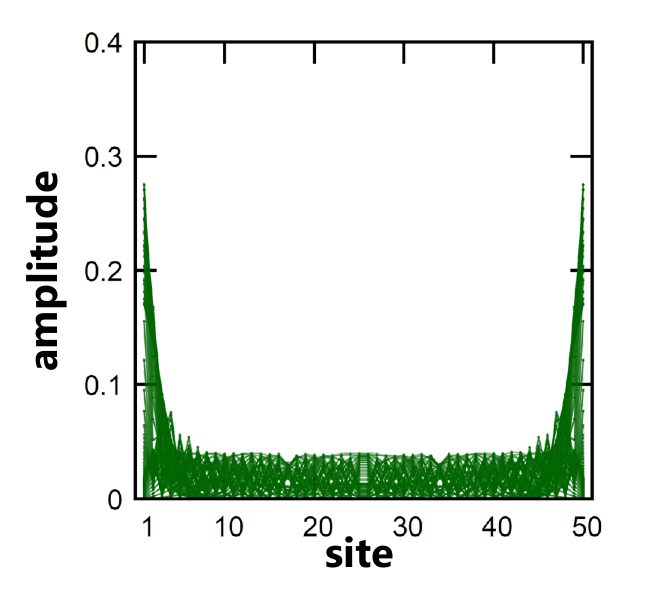} 
\caption{$\mathbb{Z}_2$ skin effect in a time-reversal-invariant topological insulator ($t=1.0$, $m=-1.6$, $\Delta=0.3$). 
The amplitude of the right eigenstates of the effective non-Hermitian Hamiltonian are shown.}	
    \label{fig: bhz}
\end{figure}

We demonstrate the $\mathbb{Z}_2$ skin effect in a time-reversal-invariant topological insulator.
As a prototypical model, we study the Bernevig-Hughes-Zhang (BHZ) model~\cite{BHZ-06} with the spin-orbit coupling, 
\begin{equation}
    H_{\mathrm{BHZ}} \left( k_x, k_y \right) = \left( t \sin{k_x}\right) \sigma_z \tau_x +
    \left( t \sin{k_y}\right) \tau_y + \left( m + t \cos{k_x}+t \cos{k_y} \right) \tau_z + \Delta \sigma_x\tau_y,
\end{equation} 
where $\sigma_i$'s and $\tau_i$'s ($i=x,y,z$) are Pauli matrices, and $t, m, \Delta \in \mathbb{R}$ are real parameters.
The BHZ model preserves time-reversal symmetry $\mathcal{T} H \left( \bm{k} \right) \mathcal{T}^{-1} = H \left( -\bm{k} \right)$ with $\mathcal{T}=\ii \sigma_y \mathcal{K}$ ($\mathcal{T}^2 = -1$) and belongs to class AII, to which the $\mathbb{Z}_2$ topological invariant is assigned.
We apply the open boundary conditions along the $x$ direction and regard the edge at $x=1$ as a system and the remaining bulk as an environment.
In Fig.~\ref{fig: bhz}, we show the right eigenstates of the effective non-Hermitian Hamiltonian under the open boundary conditions along the $y$ direction. 
We find that the Kramers pairs of the right eigenstates are localized at the opposite edges, confirming the $\mathbb{Z}_2$ skin effect.

\end{document}